\documentclass[%
 reprint,
 amsmath,amssymb,
 aps,
]{revtex4-2}

\usepackage{graphicx}
\usepackage{dcolumn}
\usepackage{bm}

\begin{document}

\title{Single-frame transmission and phase imaging using off-axis holography with undetected photons}




\affiliation{%
 Authors' institution and/or address\\
 This line break forced with \textbackslash\textbackslash
}%
\author{Emma Pearce$^{1,2,+}$}
\author{Osian Wolley$^{1,+}$}
\author{Simon P. Mekhail$^{1}$}
\author{Thomas Gregory$^{1}$}
\author{Nathan R. Gemmell$^{2}$}
\author{Rupert F. Oulton$^{2}$}
\author{Alex S. Clark$^{3}$}
\author{Chris C. Phillips$^{1}$}
\author{Miles J. Padgett$^{1,}$}
\email{miles.padgett@glasgow.ac.uk}
\affiliation{$^{1}$School of Physics and Astronomy, University of Glasgow, Glasgow, G12 8QQ, United Kingdom
}%
\affiliation{$^{2}$Blackett Laboratory, Department of Physics, Imperial College London, London, SW7 2AZ, United Kingdom
}%
\affiliation{$^{3}$Quantum Engineering Technology Labs, H. H. Wills Physics Laboratory and Department of Electrical and Electronic Engineering, University of Bristol, BS8 1FD, United Kingdom}

\date{\today}

\begin{abstract}
\noindent Imaging with undetected photons relies upon nonlinear interferometry to extract the spatial image from an infrared probe beam and reveal it in the interference pattern of an easier-to-detect visible beam. Typically, the transmission and phase images are extracted using phase-shifting techniques and combining interferograms from multiple frames. Here we show that off-axis digital holography enables reconstruction of both transmission and phase images at the infrared wavelength from a single interferogram, and hence a single frame, recorded in the visible. This eliminates the need for phase stepping and multiple acquisitions, thereby greatly reducing total measurement time for imaging with long acquisition times at low flux or enabling video-rate imaging at higher flux. With this single-frame acquisition technique, we are able to reconstruct transmission images of an object in the infrared beam with a signal-to-noise ratio of $1.78\,\pm\,0.06$ 
at 10 frames per second, and record a dynamic scene in the infrared beam at 33 frames per second.

\end{abstract}

\maketitle


\noindent Imaging and sensing with undetected photons is now a well-established and actively researched technique for probing samples in the infrared (IR) while only ever detecting visible light \cite{LemosTutorial}. Motivating this is the ability to observe highly specific molecular information by using IR light to investigate a sample's vibrational and rotational modes, with use in polymer identification for recycling \cite{Plastics}, gas sensing \cite{MIRgas}, quality control \cite{MEMS, Ceramics}, and diagnostic medicine \cite{IRcancerreview}. These applications are, however, limited by currently available IR detection technologies; cameras in the IR remain significantly more expensive, less efficient, and noisier than silicon-based detectors for visible wavelengths.

To avoid the shortcomings of IR detection, imaging with undetected photons (IUP) exploits a nonlinear interferometer to decouple the probe and detection wavelengths. Photon pairs containing a visible photon (signal) and an infrared photon (idler) are generated, typically by spontaneous parametric down-conversion (SPDC) in a nonlinear crystal. By passing the pump through the crystal twice, pairs can be generated in either pass. Pairs generated in the first pass are precisely overlapped to be indistinguishable from pairs generated in the second pass, leading to interference in the signal photon count rate \cite{ZWM1991}. Adding an object to the path of the infrared photons from the first pass introduces distinguishability. This manifests as a reduction in the amplitude of the visible interference fringes, which is proportional to the object transmissivity. Changes in phase, which are proportional to the refractive index of the object, are also observed as a phase shift in the visible interference fringes. Importantly, coincident detection is not necessary as it is only the potential to distinguish which affects the interference. The presence of an object in the idler beam can be therefore be recorded in the signal photon channel without ever detecting the infrared idler photons which interact with the object \cite{Lemos2014}. This approach has been applied to a wide variety of fields such as spectroscopy \cite{SpectroscopyPaterova,VisFTIR,NIRgratingRamelow,AgGaS2Paterova,AgGaS2Mukai}, optical coherence tomography \cite{OCTPaterova,OCTRamelow20}, hyperspectral imaging \cite{paterova2020hyperspectral}, and microscopy \cite{MicroscopyRamelow,BiologicalMicroscopy}.

The phase and transmission images are typically obtained from multiple frames acquired at different phase shifts to reconstruct the full oscillation at each pixel. Most commonly, this is achieved by a small movement of either the signal or idler mirror using a motorised stage, which we will refer to as phase-shifting, with an image taken at each position for a minimum of 3 positions \cite{CompactPaper}. This significantly affects the overall measurement rate, requiring time for the stage to move and settle as well as multiple exposure times. Capturing multiple images becomes increasingly time-consuming when long camera exposures are required due to low SPDC flux. In these types of IUP systems, it is possible to image the transmission profile directly, and therefore quickly \cite{VideoRateGrafe}, but the phase image is only obtainable by phase-shifting \cite{MicroscopyRamelow,paterova2020hyperspectral,GrafeHolography}. Alternatively, it has been shown that full phase and transmission information can be obtained in a single camera frame by splitting the detected beam into 4 regions on the sensor, each with a different phase shift introduced, although this comes at the expense of reduced intensity per pixel and many additional optics \cite{PhaseQuadrature}.

Here we demonstrate a recently proposed single-frame approach \cite{LemosTutorial} that uses off-axis digital holography. In off-axis holography, a spatial carrier frequency is introduced between the two interfering waves, referred to here as the image and reference waves \cite{Leith:62}. This is achieved by introducing an angle between the image and reference beams, typically by a combination of tilting the mirrors or beamsplitter in the interferometer and adjusting the path delay until straight tilt fringes are detected. The carrier frequency introduced separates the $+1$, $-1$, and DC diffracted orders from each other in $\mathbf{k}\text{-space}$, which allows for spatial filtering in order to select either the $+1$ or $-1$ order \cite{Cuche:00}. Reconstruction of the object field is then possible from a single interferogram. 

In off-axis holography with a conventional linear interferometer, the interference pattern detected on the camera takes the form \begin{equation}
I_{tot} = I_{im} + I_{ref} + 2\Re \left[ E_{im}(\mathbf{r}) E_{ref}^*(\mathbf{r}) \exp\left(i\mathbf{k_{tilt}}\mathbf{r}\right) \right],
 \label{eq:OffAxis}\end{equation} 
 
 \noindent where $I_{im}$ and $I_{ref}$ are the image and reference intensities, $E_{im}$ and $E_{ref}$ the complex fields of the image and reference (where the object transmission and phase is contained within the image field) and $\mathbf{k_{tilt}}$ is the relative wavevector between fields introduced as a spatial carrier frequency. Typically the non-interfering terms $I_{im}$ and $I_{ref}$ are subtracted using pre-recorded images and a Fast Fourier Transform (FFT) of the remaining terms yields the two interference terms which are separated in $\mathbf{k}\text{-space}$. Spatial filtering allows for the selection of one of these terms which, when an inverse FFT (IFFT) is performed, the result corresponds to $E_{im}(\mathbf{r}) E_{ref}^*(\mathbf{r}) \exp\left(i\mathbf{k_{tilt}}\mathbf{r}\right)$. The measurement of $I_{ref}$ allows for an estimate of $E_{ref}$ assuming a flat phase profile, and the linear phase ramp $\mathbf{k_{tilt}}$ can be calculated from the location of the centre of the term in the Fourier transform. This allows for a calculation of the field information about the object contained within $E_{im}$. Similarly, in a nonlinear interferometer for IUP in the low gain regime, the recorded intensity takes the form \begin{widetext}
\begin{equation}
I_{tot} = I_{sig,1} + I_{sig,2} + 2\Re \left[ A_{sig,1}(\mathbf{r}) A_{sig,2}(\mathbf{r}) \exp\left(i\phi + i\mathbf{k_{tilt}}\mathbf{r}\right) \left[T(\mathbf{r}) \exp\left(i\phi_{object}(\mathbf{r}) \right) \right]^{2}\right],
\label{eq:offaxiseq}
\end{equation}
\end{widetext}

\noindent where the object is described by a position-dependent transmission $T(\mathbf{r})$ and phase $e^{i\phi_{object}(\mathbf{r})}$, which is squared as the object is passed twice in the Michelson interferometer. Comparing with Eq. \ref{eq:offaxiseq}, we see that  $I_{sig,1} + I_{sig,2}$ can be determined by recording a frame when the idler beam is blocked. This can then be subtracted from the total intensity. An FFT of the subtracted image can be taken and spatial filtering performed to select one of the diffracted orders. This gives $ A_{sig,1}(\mathbf{r}) A_{sig,2}(\mathbf{r}) T^{2}(\mathbf{r}) \exp\left(i\phi + 2i\phi_{object}(\mathbf{r}) + i\mathbf{k_{tilt}}\mathbf{r}\right)$ after an IFFT. We assume that the amplitude profile of the first and second pass signal beams, $A_{sig,1}(\mathbf{r})$ and $ A_{sig,2}(\mathbf{r})$, respectively, are approximately equal, and so division by half the measured DC signal, $(I_{sig,1} + I_{sig,2})$, yields $T^{2}(\mathbf{r}) \exp\left(i\phi + 2i\phi_{object}(\mathbf{r}) + i\mathbf{k_{tilt}}\mathbf{r}\right)$. The phase ramp $\mathbf{k_{tilt}}$ is calculated in exactly the same way as in a conventional off-axis scheme. The global phase ($\phi$) can also be removed in a pre-calibration step by calculation of the phase without an object present. This allows for a calculation of the object transmission $T(\mathbf{r})$ and phase $\phi_{object}(\mathbf{r})$ from a single camera frame. 


\section*{Results}
\subsection*{Imaging System}

\begin{figure*}[ht]
\centering
\includegraphics[width=0.7\linewidth]{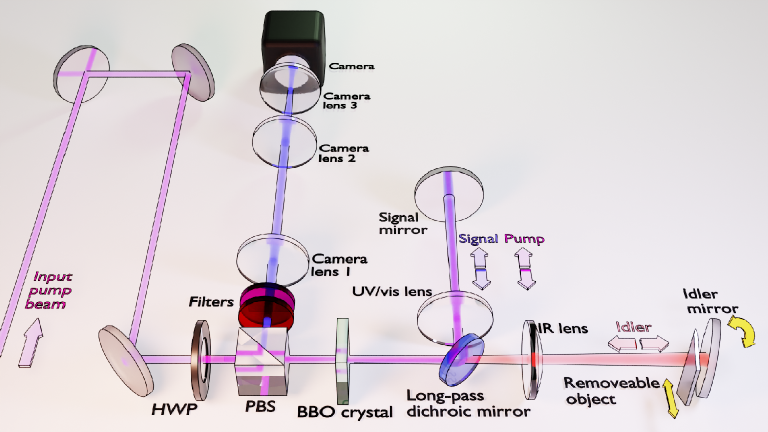}
\caption{Schematic of experimental setup for off-axis holography with undetected photons. Purple indicates the pump beam (UV, 355\,nm). Blue and red represent signal (visible, 460\,nm) and idler (SWIR, 1555\,nm) beams, respectively.}
\label{fig:SetUp}
\end{figure*}
\noindent
Figure \ref{fig:SetUp} shows the experimental setup used to perform off-axis digital holography with undetected photons. A 275\,mW continuous-wave laser at a wavelength of 355\,nm is prepared in polarization by a half-wave plate (HWP) and a polarizing beam-splitter (PBS) before pumping a 5\,mm-long beta barium borate (BBO) crystal. Signal (460\,nm) and idler (1555\,nm) pairs are produced via type-1 SPDC. The idler photons are separated by a long-pass dichroic mirror, forming the two paths of the interferometer. Signal and pump propagate together towards the signal mirror, while the idler is sent towards the idler mirror. The object to be imaged can be optionally inserted in front of the idler mirror. A lens in each arm ($f$\,=\,100\,mm) means that the mirrors sit in the Fourier plane of the crystal. All three wavelengths are then reflected and pass back through the crystal. Upon the pump's second pass, there is a probability to generate another signal-idler photon pair. The signal and idler photons are then separated from the pump at the PBS, with any residual pump and idler removed by a combination of filters. The camera is placed in a projected Fourier plane of the crystal.  The idler photons which interact with the object are never detected.


Straight fringes can be introduced by tilting the idler mirror \cite{NLI_TiltFringes,MichNLITiltFringes}, denoted by the curved yellow arrow in Fig. \ref{fig:SetUp}. Achieving this is possible because the interference pattern depends on the phase of all fields within the interferometer. This is a notable difference when compared with typical off-axis digital holography, where it is necessary for the detected beams to be tilted relative to each other. The desired fringe pattern may also be achieved by adjusting the signal mirror in this system, however, we choose to optimize the signal mirror position for maximum SPDC intensity at the camera. We then adjust the idler mirror, which affects only the interference and not the intensity (in the low-gain regime) as it is not directly detected. Adjusting the signal mirror would also affect the pump beam returning to the BBO and hence alter the phase-matching.

\begin{figure*}[ht]
\centering
\includegraphics[width=0.95\linewidth]{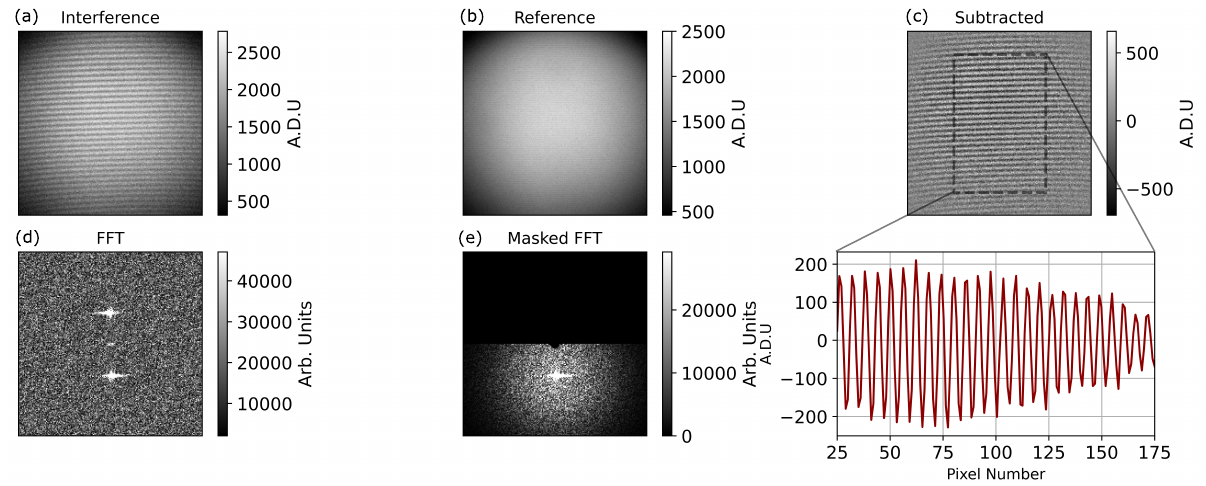}
\caption{Digital off-axis holography acquisition and analysis. Example frames showing (a) interference, (b) the reference beam acquired by blocking the infrared path, and (c) interference with the reference subtracted. Raw camera frames are shown in Analog-to-Digital Units (A.D.U.) A vertical cut averaged over the rows (zoomed inset) shows $\sim25$ fringes with a maximum  $22.8\,\pm\,0.9\%$ contrast. Also shown is (d) the FFT of the subtracted image and (e) the mask used in the image reconstruction.}
\label{fig:FringeCont}
\end{figure*}

In order to perform the spatial filtering required to isolate the +1 diffracted order from the DC and $-1$ orders, enough tilt fringes must be introduced such that the $+1$ order is sufficiently separated in $\mathbf{k}\text{-space}$, and that the filtering applied does not compromise the resolution of the imaging system by removing high spatial frequencies from the image. However, as with a conventional interferometer, introducing more tilt fringes worsens the fringe contrast as the beams walk off from each other. Aligning the system then becomes a balance between a reasonable fringe contrast such that the signal-to-noise ratio (SNR) of images is preserved and enough tilt fringes in the field of view to maintain image resolution. Figure. \ref{fig:FringeCont} shows the interference pattern and reference subtraction used to perform off-axis holography. We align the system such that we have $\sim30$ fringes in the field of view (FOV), measured to be $12.47\,\pm\,0.03 \text{mm}$. This results in a maximum fringe contrast of $22.8\,\pm\,0.9\%$ as measured on $1000$ reference-subtracted images, defined as $(I_{max}-I_{min})/2A_{sig,1}A_{sig,2}$. This gives a maximum measurable transmission of $0.48 \pm 0.01$. A Gaussian filter is then applied, centred around $+1$ diffraction order. Subtraction of the reference removes the majority of the DC component, however we apply a mask to the central $\sim\,5$ pixels to remove DC component that remains. Figure \ref{fig:FringeCont} shows an example of an unmasked and masked FFT of the interference pattern used in the image reconstruction.

\subsection*{Transmission and Phase Imaging}

\begin{figure*}[ht]
\centering
\includegraphics[width=\linewidth]{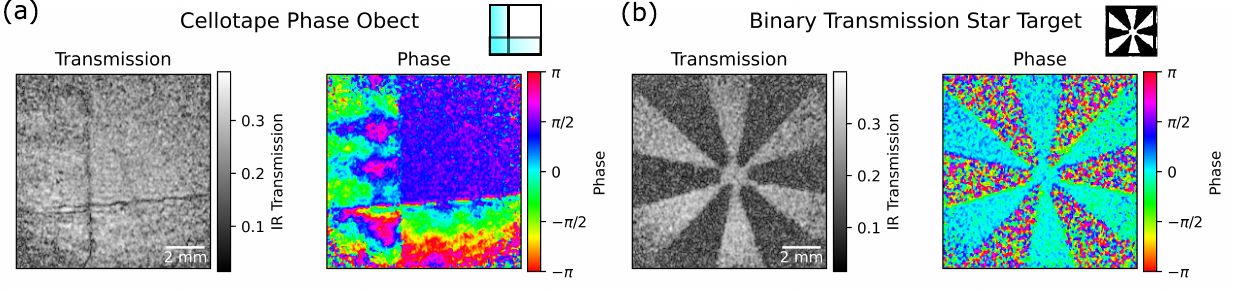}
\caption{Single-frame complex image reconstruction of (a) a phase object (overlapping pieces of cellotape) and  (b) a binary transmission object (3D printed star). Objects measured over a $12.47\,\pm\,0.03 \,\text{mm}$ field of view. Note the phase measured here is twice the optical phase thickness of the object due to the double pass.}
\label{fig:IntPhaseIms}
\end{figure*}
\noindent
Figure \ref{fig:IntPhaseIms} shows transmission and phase images of a binary transmission star-shaped target placed in the idler arm acquired with the imaging system. Both the intensity and the phase are reconstructed from a single interference pattern, which were acquired at a rate of $10\,\text{Hz}$. In order to properly demonstrate phase imaging, an example of a transmissive `phase object' made from two pieces of overlapping cellotape is also shown. From the phase images, variation in the optical thickness in the cellotape can be seen which is not present in the transmission image. Note that, in the phase images, a pre-calibrated reference phase has been subtracted in order to correct for the global phase term present in the object field and no unwrapping is currently performed. Figure \ref{fig:IntPhaseIms} demonstrates that by using off-axis holography, it is possible to reconstruct transmission and phase images from a single interference pattern in a nonlinear interferometer where the light detected has never interacted with the object.

\begin{figure*}[ht]
\centering
\includegraphics[width=0.7\linewidth]{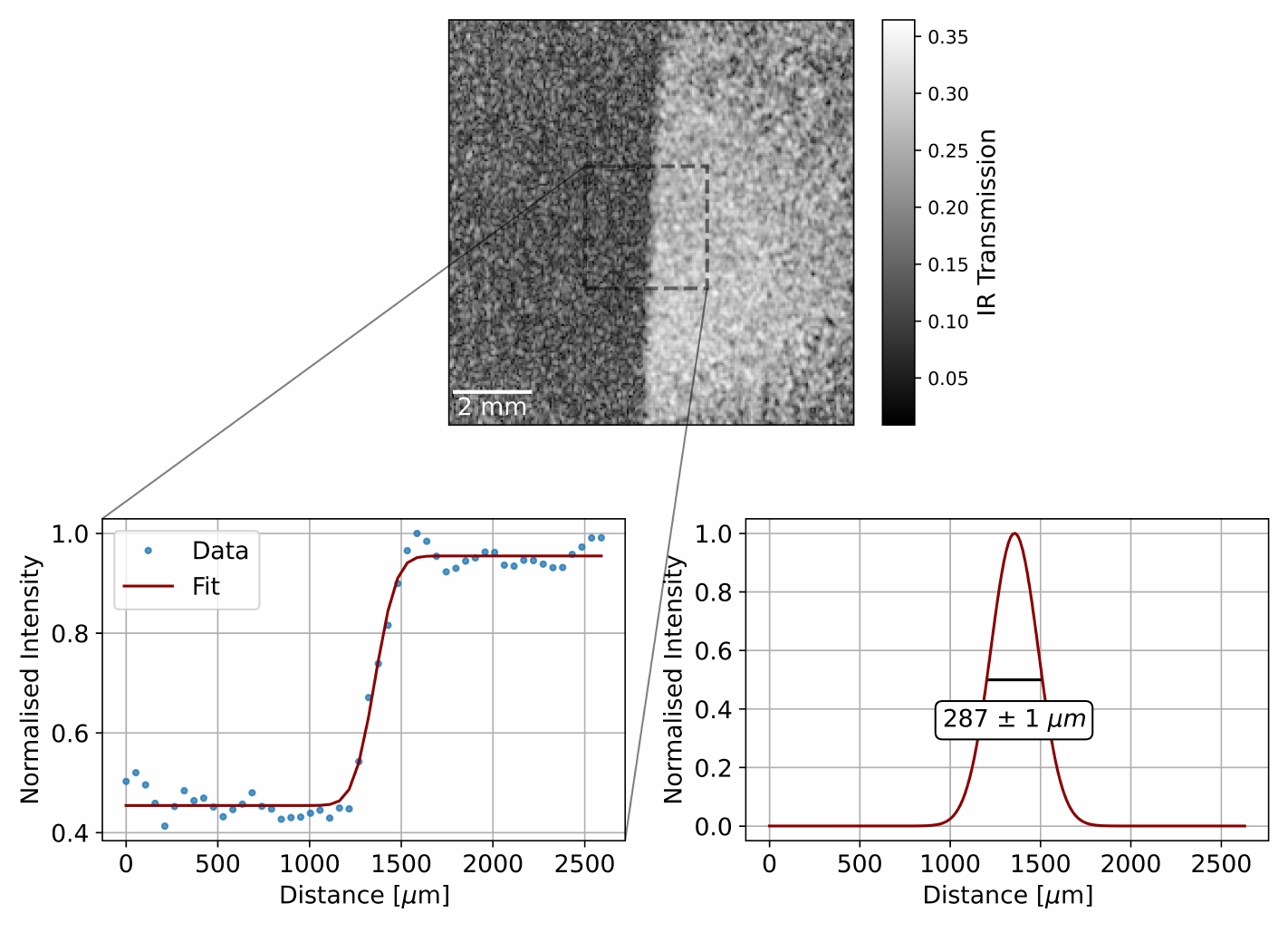}
\caption{Reconstructed image of a slanted edge placed in the idler arm, with the region used to measure the edge response of the system indicated. The tilt of the edge was corrected for and the edge response fitted with a Gaussian error function (zoomed inset). The derivative of the error function is a Gaussian, the FWHM of which is used to calculate the point spread function. This was repeated for 1000 frames with the mean value calculated and error stated as the standard error on the mean.}
\label{fig:EdgeTest}
\end{figure*}

In order to assess the performance of the imaging system, we measure the SNR and resolution using a slanted edge placed in the idler beam. To measure the resolution of the system, a Gaussian error function is fitted to a section of the edge response as indicated in Fig. \ref{fig:EdgeTest}. The derivative of the error function gives a Gaussian function from which the point spread function (PSF) of the system can be calculated as the full width at half maximum (FWHM). This calculation was averaged over $1000$ frames giving a PSF of $287\,\pm\,1 \,\mu \text{m}$. With the FOV of the system measured as $12.47\,\pm\,0.03 \, \text{mm}$, this gives an approximation of the total number of modes for the system as $1890\,\pm\,20$. In the far-field \cite{MicroscopyRamelow}, the FOV is calculated as $\text{FOV} = 2f\theta_{i}/M$ where $f$ is the focal length of the lens matching the far-field of the crystal to the object, $\theta_{i}$ is the emission angle of the idler photons and $M$ the magnification factor due to the difference in wavelength of signal and idler photons \cite{ResolutionMomentum}. The emission angle depends on the phase-matching conditions in the crystal, which we model to account for the broadband nature of the down-converted photons \cite{Boeuf2000}. From this calculation, we would expect a FOV of $\sim$\,12.97\,mm, accounting for the 5\,nm bandpass filter used in the experiment. The resolution is calculated as \begin{equation}
    \delta x_{corr} = \frac{\sqrt{2\ln{2}}f\lambda_{i}}{\pi w_{p}M},
\end{equation} where $\lambda_{i}$ is the idler wavelength and $w_{p}$ the pump beam waist. For the imaging system here, this gives a resolution of $\sim$\,406\,$\mu$m. The total number of spatial modes can then be calculated as \begin{equation}
    m_{2D} = \left ( \frac{\text{FOV}}{\delta x_{corr}} \right) ^2, 
\end{equation} giving a total number of modes as $\sim$\,1000. The difference between the theoretical and measured values for the resolution are likely due to error in the value of the pump beam waist, which was taken from the manufacturer specification sheet as $1.0\,\pm\,0.2 \, \text{mm}$. The difference in the FOV values can be explained by slight discrepancy in the angle of the BBO compared to the nominal value used in calculations and the broadband nature of the down-conversion. The angle of emission, and hence the resulting FOV, are sensitive to the centre wavelength and bandwidth of the bandpass filter placed before the camera. As we slightly tilt the filter to optimise the detection of the signal, it is difficult to precisely estimate the centre wavelength and bandwidth of the down-conversion. Here, the difference between the measured and theoretical FOV corresponds to only a $0.04^{\circ}$ difference in emission angle and the difference in measured and theoretical resolution values corresponds to a difference in pump waist of $0.4\,\text{mm}$. As the predicted resolution does not assume anything about the method of reconstruction, we conclude that we see no deterioration of resolution in the imaging system by using the off-axis approach.


We measure the SNR of the reconstructed transmission images by masking the bright and dark regions of the image and average over $1000$ images. This gives a SNR value of $1.78\,\pm\,0.06$. We note that, as mentioned previously, there is a trade off between SNR and resolution with the size of the mask applied in Fourier space, with higher SNR attainable with a smaller mask but at the cost of a reduction in the spatial resolution. 


\subsection*{Video Rate Imaging}
\begin{figure*}[ht]
\centering
\includegraphics[width=\linewidth]{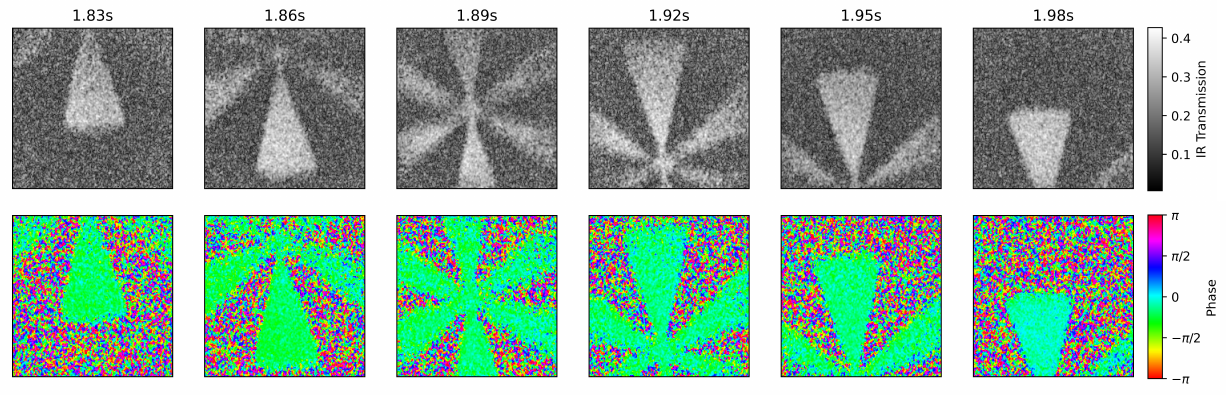}
\caption{Reconstructed transmission and phase images from selected frames from a 33 frames per second video of a moving binary transmission target.}
\label{fig:VideoRate}
\end{figure*}

\noindent
The single-frame nature and Fourier transform-based analysis of this approach means that not only can images be recorded at video rate, but the transmission and phase information can also be extracted at this speed. The analysis can therefore be easily integrated into a graphical user interface. Figure~\ref{fig:VideoRate} shows a sequence of images from a video of the binary transmission target as it is moved across the IR path (Supplementary Video), captured and analysed at 33 frames per second (fps).

\section*{Discussion}
\noindent
Most IUP implementations use quasi-phase-matching in periodically-poled nonlinear crystals for photon pair generation, offering strong nonlinearities and tailored photon pair wavelengths. However, the crystal aperture is often only 1\,-\,2\,mm due to the poling process. When significant tilt is introduced by the mirror in far-field in the IR path, the returning light will be laterally displaced at the crystal and potentially clipped by the crystal aperture. Here, the bulk BBO crystal's large transverse dimensions (10\,mm $\times$ 10\,mm) allow significant tilt to be introduced in the IR path without concern. This effect is evident here when tilting the signal mirror instead, as the first and second pass of the pump are visibly separated at the crystal when many fringes are present in the interferogram. The geometric constraints of the crystal aperture should therefore be considered when applying this technique to IUP systems. 

The main drawback in performing off-axis holography for imaging with undetected photons when compared with phase-shifting is the trade-off between SNR and resolution introduced by the tilt fringes themselves and the filtering process. Whilst in principle it is possible to perform off-axis holography without loss in resolution, obtaining a sufficient angle between interfering beams is a challenge within conventional holography and can be a further challenge in a nonlinear interferometer due to the short coherence lengths and often low visibility. Whilst an off-axis approach is unlikely to produce images with comparably high SNR and resolution as phase-shifting, it does have a significant advantage in the speed of both image acquisition and image processing, as well as experimental complexity. 

Digital holography is itself a large field of research, with many approaches to optimising image reconstruction \cite{Luo:23,DL,AdvancesDHI}. While we use a simple approach to image reconstruction here, there is much scope for future work exploring more complicated protocols which may address some of the resolution and SNR issues discussed above \cite{khare_single_2013}. The demonstration of imaging with undetected photons in an off-axis configuration allows for the extension of many techniques within digital holography literature to be explored at previously challenging wavelengths, for example off-axis spatial multiplexing \cite{Shaked:20} and quantitative phase imaging techniques for disease identification \cite{kastl:2017}. 


In conclusion, we have applied off-axis digital holography to imaging with undetected photons for the first time. We are able to reconstruct both intensity and phase images of binary and transmissive objects from a single interference image. This facilitates video-rate imaging of dynamic scenes within the infared path, which we demonstrate at 33\,fps. Our demonstration helps extend the potential applications of imaging with undetected photon schemes to scenarios where single-shot imaging and frame rate are important considerations, such as the monitoring dynamic biological processes and rapid industrial assessments of materials \cite{Kemper2008,Roadmap}.

\section*{Methods}
\noindent
In order to generate signal and idler pairs, a $355\,\text{nm}$ wavelength continuous-wave laser (Coherent CX355-250 CW) with nominal beam waist $1.0\,\pm\,0.2 \,\text{mm}$ is prepared in polarisation and pumps a $10\,\times\,10\,\times\,5$\,mm$^{3}$  BBO crystal. The crystal is nominally cut to produce non-degenerate photon pairs at $460\,\text{nm}$ (signal) and $1555\,\text{nm}$ (idler) via type-1 SPDC. The idler photons are separated by a long-pass dichroic mirror (Chroma T425lpxr). Both mirrors are placed in the far-field of the crystal by use of $f$\,=\,100\,mm lenses. A 5\,nm bandpass filter (Semrock BrightLine FF01-461/5-25) centred at 461\,nm is placed after a PBS which directs photon pairs to the camera. Camera lens 1 ($f$\,=\,100\,mm) also places the camera in the far-field of the crystal, where we use a relay, camera lenses $2$ ($f$\,=\,120\,mm) and $3$ ($f$\,=\,35\,mm), to demagnify by a factor of 3.4 onto the camera. The camera used is a Hamamatsu ORCA-Quest (C15550-20UP), run in ultra-quiet scan mode with a pixel size of 4.6\,$\mu$m.  
\newline

\section*{Acknowledgements}
\noindent
We acknowledge funding from the UK National Quantum Hub for Imaging (QUANTIC, No. EP/T00097X/1) and the Royal Society (No. UF160475). This work was funded by
the UK EPSRC (QuantIC EP/M01326X/1). O.W. acknowledges the financial
support from the EPSRC (QuantIC EP/R513222/1). S.P.M. acknowledges the
EPSRC (QuantIC EP/M01326X/1) and the Leverhulme Trust. 
T.G. acknowledges the financial support from the EPSRC (EP/SO19472/1).
M.J.P. acknowledges the financial support from the Royal Society
(RSRP/R1/211013P).

\section*{Author contributions statement}
\noindent
E.P., O.W., S.P.M., T.G., N.R.G., and M.J.P. devised the project methodology. E.P. and O.W. conducted the experiments and data analysis. R.F.O., A.S.C., C.C.P, and M.J.P. provided supervision. All authors reviewed the manuscript.

\section*{Additional information}
\noindent
\textbf{Competing interests:} The authors declare no competing interests.
\section*{Data Availability}
\textbf{Data Availability Statement:} \noindent All data used to produce the figures in this work will be made available upon request. 
\bibliography{sample}

\end{document}